\documentclass[aip,apl,reprint]{revtex4-1}
\usepackage{bm}
\usepackage{amsmath,amssymb}
\usepackage{here}
\usepackage[]{graphicx,color}
\usepackage{CJKutf8}
\usepackage{xspace}

\definecolor{darkgreen}{rgb}{0,0.5,0}
\definecolor{grey20}{rgb}{0.80,0.80,0.80}

\begin{document}

\title{Cerium as a possible stabilizer of ThMn$_{12}$-type iron-based compounds:
  A first-principles study}

\affiliation{Research Center for Computational Design of Advanced Functional Materials, 
  National Institute of Advanced Industrial Science and Technology, 
  Tsukuba, Ibaraki 305-8568, Japan}
\affiliation{Elements Strategy Initiative Center for Magnetic Materials, 
  National Institute for Materials Science, Tsukuba, Ibaraki 305-0047, Japan}
\affiliation{Institute of Materials and Systems for Sustainability, 
  Nagoya University,
  Nagoya, Aichi 464-8601, Japan}

\author{Yosuke Harashima}
\affiliation{Research Center for Computational Design of Advanced Functional Materials, 
  National Institute of Advanced Industrial Science and Technology, 
  Tsukuba, Ibaraki 305-8568, Japan}
\affiliation{Elements Strategy Initiative Center for Magnetic Materials, 
  National Institute for Materials Science, Tsukuba, Ibaraki 305-0047, Japan}
\affiliation{Institute of Materials and Systems for Sustainability, 
  Nagoya University,
  Nagoya, Aichi 464-8601, Japan}

\author{Taro Fukazawa}
\affiliation{Research Center for Computational Design of Advanced Functional Materials, 
  National Institute of Advanced Industrial Science and Technology, 
  Tsukuba, Ibaraki 305-8568, Japan}
\affiliation{Elements Strategy Initiative Center for Magnetic Materials, 
  National Institute for Materials Science, Tsukuba, Ibaraki 305-0047, Japan}

\author{Takashi Miyake}
\affiliation{Research Center for Computational Design of Advanced Functional Materials, 
  National Institute of Advanced Industrial Science and Technology, 
  Tsukuba, Ibaraki 305-8568, Japan}
\affiliation{Elements Strategy Initiative Center for Magnetic Materials, 
  National Institute for Materials Science, Tsukuba, Ibaraki 305-0047, Japan}



\date{\today}

\begin{abstract}
  The structural stability of CeFe$_{12}$ is investigated by using first-principles calculation. 
  The formation energies of CeFe$_{12}$ relative to 
  the Ce$_{2}$Fe$_{17}$ + bcc-Fe phase and to the CeFe$_{2}$ + bcc-Fe phase
  are calculated with the assumptions of trivalency and tetravalency for Ce.
  Those values are compared with corresponding results in $R$Fe$_{12}$ for $R=$ Nd, Sm, and Zr.
  Our results suggest that the tetravalent Ce is a promising stabilizer of the ThMn$_{12}$ structure.
  We also show that the stabilizing effect of an element depends as much on the valency as on the size of the $R$ element
  by investigating $R$Fe$_{12}$ where $R$ is assumed to have a hypothetical valency
  on the basis of first-principles calculation.
\end{abstract}

\maketitle

The permanent magnet is one of the most important materials in industries.
The range of its application is widespread: the motors in electric vehicles, wind turbine generators, hard disks, and so on.
Currently, the most widely used high-performance permanent magnet is NdFeB-based magnets, the main phase of which is 
 Nd$_{2}$Fe$_{14}$B.
Performance of a permanent magnet is
determined by the intrinsic properties of the main phase, 
and by the microstructure formed by the subphases.
Intensive research over three decades has improved
the performance of the NdFeB-based magnet
by controlling its microstructure.
However, the performance seems to reach near to its theoretical limit 
today,
hence a new magnet compound for the main phase is required. 

Rare-earth compounds with the ThMn$_{12}$ structure have been regarded as
potential strong magnet compounds 
since iron-rich phases, such as SmFe$_{11}$Ti, were synthesized 
in the late 1980's.~\cite{OhTaOsSh1988,Bu1988,YaKoSuGuCh1988} 
Compounds with high iron content, which leads to large magnetization,
has been investigated intensely these days.
Hirayama et al. have succeeded in synthesizing NdFe$_{12}$N~\cite{HiTaHiHo2015}, 
stimulated by a theoretical work on NdFe$_{12}$ and NdFe$_{12}$N~\cite{MiTeHaKiIs2014}.
They measured intrinsic magnetic properties, and found that 
its saturation magnetization, anisotropy field and Curie temperature 
are higher than those of Nd$_{2}$Fe$_{14}$B. 

The thermodynamic stability has also been one of the central issues
in exploration of high-performance ThMn$_{12}$ compounds.
Unfortunately,
NdFe$_{12}$N (and its mother compound, NdFe$_{12}$) exists only as a film due to its instability.
Introducing another element can stabilize the ThMn$_{12}$ structure,
and it affects
the magnetic properties significantly. 
For example,
it has already been known in the 1980's that
substitution of Ti for one of 12 Fe sites improves the stability,
but it significantly reduces
the saturation magnetization.~\cite{Fr1958,VeBoZhBu1988,MiTeHaKiIs2014,HaTeKiIsMi2016} 
Partial substitution of Zr for the rare-earth element
has eagerly investigated recently with the expectation of improvement in stability
and of small harm in its magnetism.~\cite{SuKuUrKoSaWaKiKaMa2014,SuKuUrKoSaWaYaKaMa2016,KuSuUrKoSaYaKaMa2016,SaSuKuUrKoYaKaMa2016,GjPsDeNiHa2016,GaHa2016,GaHa2017,KoSuKuUrSaYaShKaMa2017} 
In a previous work, we reported first-principle calculation of $R$Fe$_{12}$ for 14 $R$ elements:
12 rare-earth elements (La, Pr, Nd, Sm, Gd, Dy, Ho, Er, Tm, Lu, Y, Sc) 
and 2 group IV elements (Zr, Hf).~\cite{HaFuKiMi2018} 
On the basis of 
the calculated formation energy relative to the $R_{2}$Fe$_{17}$ + bcc-Fe phase,
we showed
that some elements has potential in stabilizing the ThMn$_{12}$ structure.

\begin{figure}[t]
  \includegraphics[width=\hsize]{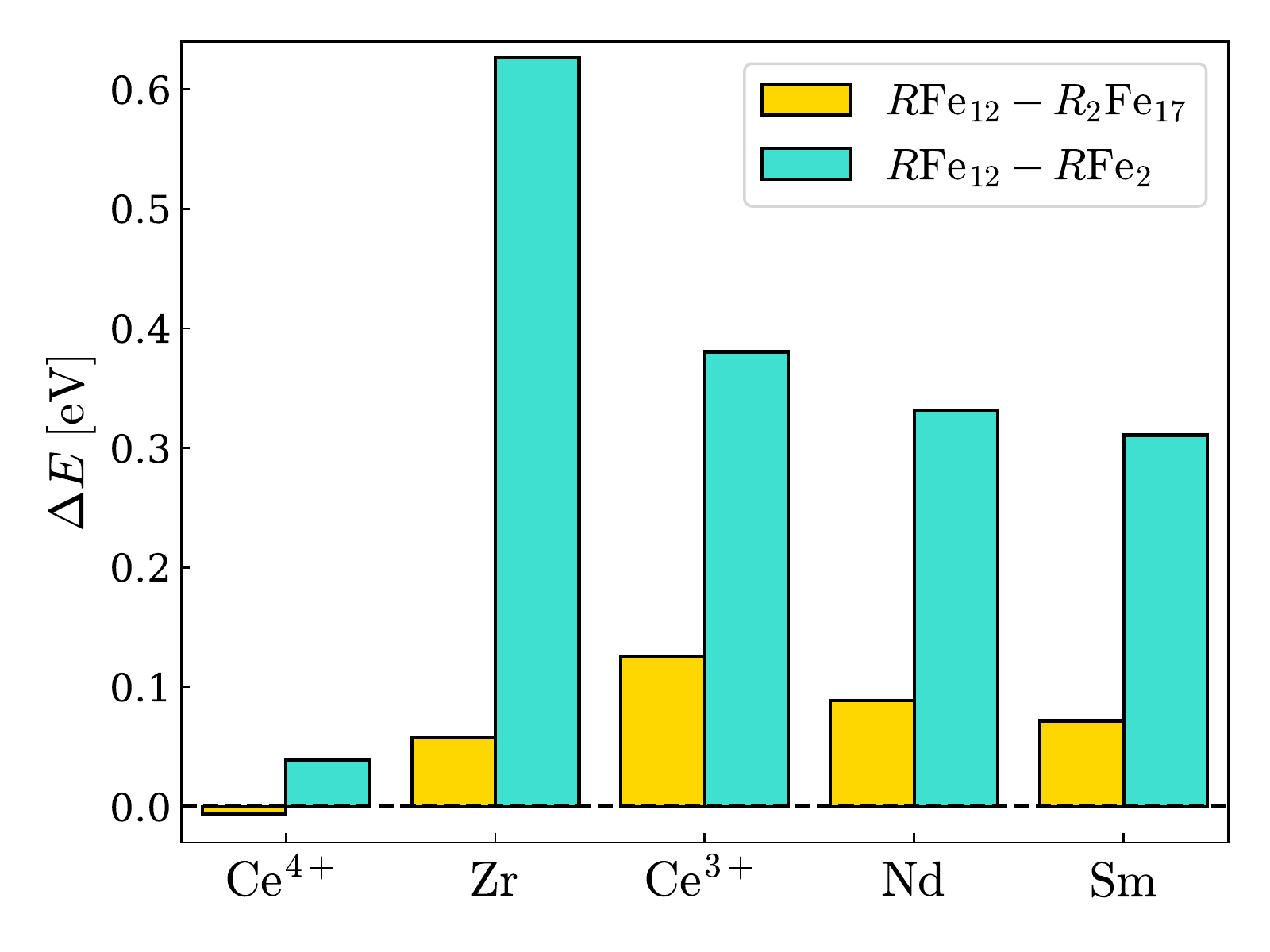}
  \caption{(Color online) The formation energy of $R$Fe$_{12}$ forming from $R_{2}$Fe$_{17}$ $+$ bcc-Fe (yellow bars)
    and from $R$Fe$_{2}$ $+$ bcc-Fe (green bars)
    for $R=$ Ce$^{3+}$, Ce$^{4+}$, Nd, Sm, and Zr are shown.
    They are described in electron volts per formula unit of $R$Fe$_{12}$.}
  \label{fig:formationenergy_1-12_2-17_1-2}
\end{figure}

In the present work, we investigate potential of Ce as a stabilizer of 
the ThMn$_{12}$ structure.
Cerium is an abundant element, and often called as a ``free'' rare-earth because 
it is much more inexpensive than the other rare-earths.
It has been attempted to replace other rare-earth elements in the ThMn$_{12}$-type structure
with Ce.~\cite{GaMaBaSaHa2016,GaHa2016,MaSaScGaBaHa2018}
Cerium is usually tetravalent~\cite{BuWi1970} or in a mixed valence state~\cite{CoAlMiBu1993} in rare-earth--iron compounds.
This is quite distinct from other rare-earth elements:
Most of rare-earth elements are trivalent, and 
some of them, e.g. Eu and Yb, are divalent, 
but none of them are tetravalent in $R$--Fe compounds. 
Although the magnetism of Ce-doped ThMn$_{12}$-type
compounds has been studied well especially in experiments,
there is no study on the potential of Ce as a stabilizer
of the ThMn$_{12}$ structure to the best of our knowledge.

To investigate the possibility, we perform first-principles calculation 
of the formation energy of CeFe$_{12}$, and 
compare the result for other $R$Fe$_{12}$ compounds. 
We analyze the result in terms of the atomic radius and valency of the $R$ element,
referring to the results of our previous study.
We also provide new calculational results
in which $R$ is assumed to be tetravalent 
for the analysis.

We use QMAS, a first-principles calculation code, which is constructed
on the basis of density functional theory~\cite{HoKo1964,KoSh1965} 
and the projector augmented-wave method~\cite{Bl1994,KrJo1999}. 
The exchange-correlation functional is 
approximated by the generalized gradient approximation of the Perdew--Burke--Ernzerhof (PBE) formula~\cite{PeBuEr1996}. 
8 $\times$ 8 $\times$ 8 $k$-points are sampled, and 
the cutoff energy for the plane wave basis is set to 40.0 Ry.
The 4$f$ electrons of rare-earth elements are treated as open-core states.
Occupation of those open-core states satisfies Hund's first rule: 
The minority spin channel is first occupied, then, if electrons remain, the majority spin channel is occupied.
For detailed discussion of the reliability of the open-core treatment in
calculation of the formation energy, we refer readers to
Ref.~\onlinecite{HaFuKiMi2018}.


\begin{table}
  \caption{
    The calculated lattice constants ($a$ and $c$),
    volume per formula unit ($V$),
    and inner coordinates of 
    CeFe$_{12}$ are shown for the trivalent and tetravalent Ce.
     They have the tetragonal ThMn$_{12}$-type structure
    (space group $I4/mmm$, No.~139).}
  \vspace{5pt}
  \begin{tabular}{lcc}
    \hline \hline
                    & Ce$^{3+}$Fe$_{12}$   & Ce$^{4+}$Fe$_{12}$ \\
    \hline
    $a$ [\AA]       & 8.579           & 8.480 \\
    $c$ [\AA]       & 4.678           & 4.673 \\
    $V$ [\AA$^{3}$] & 172.1           & 168.0 \\
    Ce (2$a$)       & (0,\;0,\;0)      & (0,\;0,\;0) \\
    Fe (8$f$)       & (1/4,\;1/4,\;1/4) & (1/4,\;1/4,\;1/4) \\
    Fe (8$i$)       & (0.360,\;0,\;0)   & (0.360,\;0,\;0) \\
    Fe (8$j$)       & (0.265,\;1/2,\;0) & (0.270,\;1/2,\;0) \\
    \hline \hline
  \end{tabular}
  \label{table:latticeproperty_1-12}
\end{table}
\begin{table}
\caption{The lattice constants ($a$ and $c$),
    volume per formula unit ($V$),
    and inner coordinates of 
    Ce$_{2}$Fe$_{17}$ are shown for the trivalent and tetravalent Ce.
    They have the rhombohedral Th$_{2}$Zn$_{17}$-type structure
    (space group $R\bar{3}m$, No.~166).}
  \vspace{5pt}
  \begin{tabular}{lccc}
    \hline \hline
                    & Ce$^{3+}_{2}$Fe$_{17}$   & Ce$^{4+}_{2}$Fe$_{17}$ & Exp.\cite{KrFeHeEzJo1996} \\
    \hline
    $a$ [\AA]       & 8.605                  & 8.459 & 8.496 \\
    $c$ [\AA]       & 12.557                 & 12.513 & 12.414 \\
    $V$ [\AA$^{3}$] & 268.4                  & 258.5 & 258.7 \\
    Ce (6$c$)       & (0,\;0,\;0.341)          & (0,\;0,\;0.342) & --- \\
    Fe (18$f$)      & (0.287,\;0,\;0)          & (0.294,\;0,\;0) & ---\\
    Fe (18$h$)      & (0.502,\;-0.502,\;0.158) & (0.500,\;-0.500,\;0.158) & --- \\
    Fe (9$d$)       & (1/2,\;0,\;1/2)          & (1/2,\;0,\;1/2) & --- \\
    Fe (6$c$)       & (0,\;0,\;0.096)          & (0,\;0,\;0.096) & --- \\
    \hline \hline
  \end{tabular}
  \label{table:latticeproperty_2-17}
\end{table}
\begin{table}
\caption{The lattice constants ($a$) and
    volume per formula unit ($V$) of CeFe$_{2}$
    are shown for the trivalent and tetravalent Ce.
    They have the cubic MgCu$_{2}$-type structure
    (space group $Fd\bar{3}m$, No.~227).}
  \vspace{5pt}
  \begin{tabular}{lccc}
    \hline \hline
                    & Ce$^{3+}$Fe$_{2}$   & Ce$^{4+}$Fe$_{2}$ &Exp.\cite{DuHi1995} \\
    \hline
    $a$ [\AA]       & 7.501              & 7.361 & 7.302 \\
    $V$ [\AA$^{3}$] & 105.5               & 99.7 & 97.3 \\
    Ce (8$a$)       & (3/8,\;3/8,\;3/8)   & (3/8,\;3/8,\;3/8) & ---\\
    Fe (16$d$)      & (0,\;0,\;0)          & (0,\;0,\;0) & ---\\
    \hline \hline
  \end{tabular}
  \label{table:latticeproperty_1-2}
\end{table}

\begin{figure}[t]
  \includegraphics[width=\hsize]{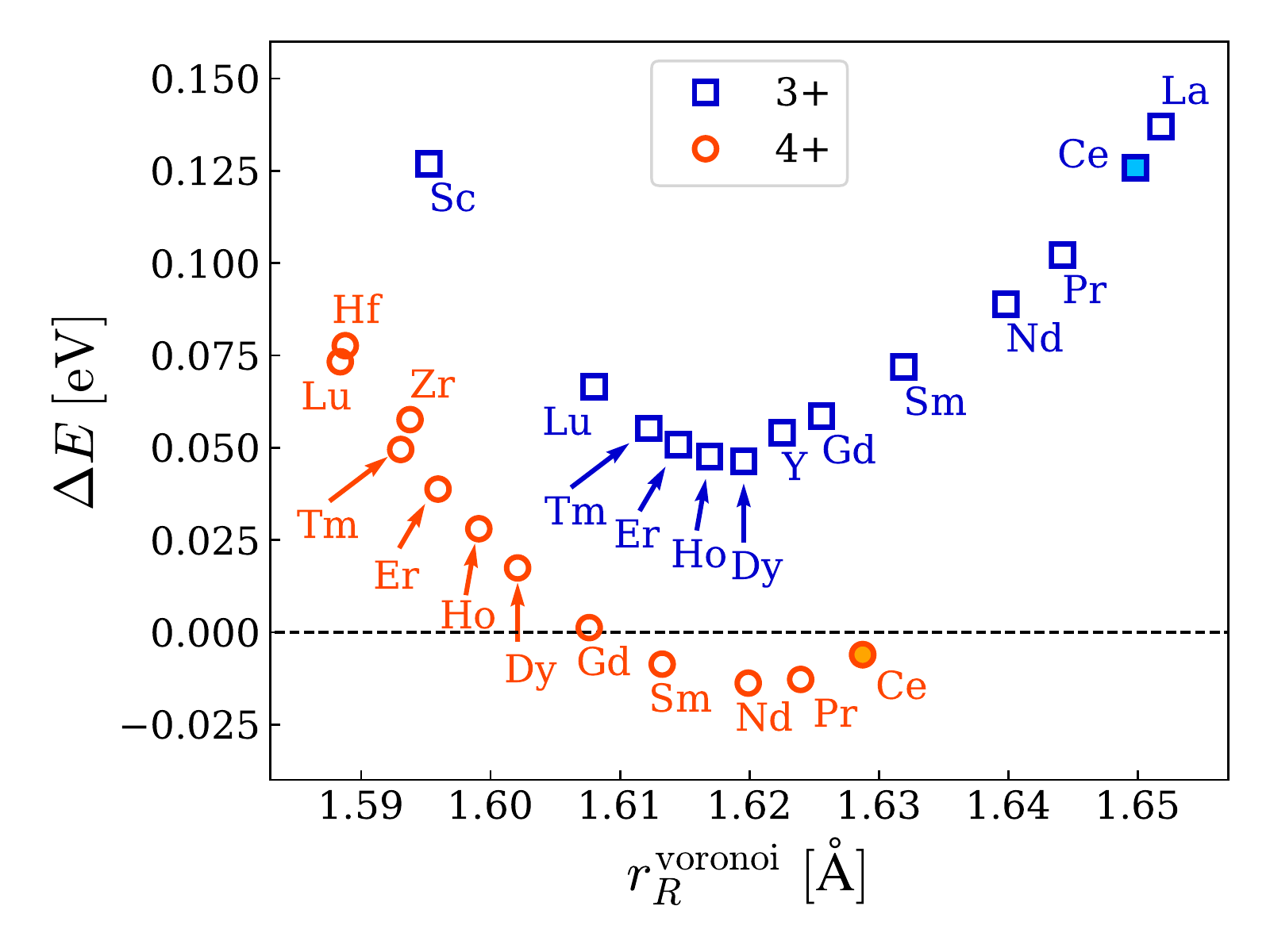}
  \caption{(Color online) The formation energy defined as the $R$Fe$_{12}$ forming from $R_{2}$Fe$_{17}$ and bcc-Fe 
    [see Eq.~\eqref{eq:formationenergy} for the definition, where Ce should be read as $R$]
    as a function of the atomic radius, $r_{R}^{\text{voronoi}}$ (see the text for its definition). 
    Blue squares are for $R^{3+}$ elements; red circles are for $R^{4+}$ elements.}
  \label{fig:formationenergy_1-12_2-17}
\end{figure}

First, we show the lattice parameters for CeFe$_{12}$
obtained by the numerical optimization, and also those for Ce$_{2}$Fe$_{17}$ and CeFe$_{2}$
for comparison.
Table~\ref{table:latticeproperty_1-12} shows the optimized structures of CeFe$_{12}$ 
calculated with 
the assumption of the Ce$^{3+}$ and Ce$^{4+}$ valency,
where the number of occupied 4$f$ electrons
are assumed to be one and zero, respectively;
Table~\ref{table:latticeproperty_2-17} shows
the result for the rhombohedral Ce$_{2}$Fe$_{17}$; 
Table~\ref{table:latticeproperty_1-2} shows
the result for the cubic CeFe$_{2}$. 

In all phases, the volume is significantly smaller with Ce$^{4+}$ than with Ce$^{3+}$. 
It can be attributed to the additional delocalized electron in the Ce$^{4+}$ systems
that is fixed to an $f$-state in the Ce$^{3+}$ systems.
The extended electron
contributes to cohesion, and makes
the difference in volume.
The experimental lattice parameters of $a=$ 8.496 \AA\; and $c=$ 12.414 \AA\; have been reported for Ce$_{2}$Fe$_{17}$~\cite{KrFeHeEzJo1996}
(although the structure is annotated to be `hexagonal' in the literature, we take its Bravais lattice as rhombohedral, which has a hexagonal
conventional cell, because the given parameters are unreasonably different from other compounds having a hexagonal Bravais lattice).
The experimental value of $a=$ 7.302 \AA\; for CeFe$_{2}$ have also been reported.~\cite{DuHi1995}
Our calculated lattice constant with Ce$^{4+}$ agrees well with the experimental values
while those with Ce$^{3+}$ deviate significantly from the experimental values.
Gabay et al. have reported 
that the volume of CeFe$_9$Co$_2$Ti having the ThMn$_{12}$ structure is slightly smaller than
that of SmFe$_9$Co$_2$Ti. \cite{GaMaBaSaHa2016}
This tendency agrees with our calculation of CeFe$_{12}$ and SmFe$_{12}$ ($a=$ 8.497 \AA\; and $c=$ 4.687 \AA \cite{HaFuKiMi2018}) when 
the tetravalency of Ce is assumed.


Next, we calculate the formation energy of CeFe$_{12}$. 
We take Ce$_{2}$Fe$_{17}$ + bcc-Fe as a
reference system
because the closest compound to CeFe$_{12}$ is Ce$_{2}$Fe$_{17}$ 
in the phase diagram. 
The definition of the formation energy is
\begin{equation}
  \Delta E \equiv E[\text{CeFe}_{12}] - \left(\dfrac{1}{2}E[\text{Ce}_{2}\text{Fe}_{17}] + \dfrac{7}{2}E[\text{Fe}]\right),
  \label{eq:formationenergy}
\end{equation}
where $E[\cdot]$ denotes the total energy per chemical formula of the system in the bracket.
This can be also expressed in terms of the formation energy of CeFe$_{12}$ and Ce$_{2}$Fe$_{17}$ from simple 
substances by $\Delta E = \Delta E^{\text{CeFe}_{12}} - \Delta E^{\text{Ce}_{2}\text{Fe}_{17}}$
where they are defined as
\begin{align}
  & \Delta E^{\text{CeFe}_{12}} \equiv E[\text{CeFe}_{12}] - \left(E[\text{Ce}] + 12E[\text{Fe}]\right),
  \\
  & \Delta E^{\text{Ce}_{2}\text{Fe}_{17}} \equiv \dfrac{1}{2} E[\text{Ce}_{2}\text{Fe}_{17}] - \left(E[\text{Ce}] + \dfrac{17}{2}E[\text{Fe}]\right).
\end{align}
We consider both the Ce$^{3+}$ and Ce$^{4+}$ valency in the following,
but the valency of Ce in CeFe$_{12}$ and Ce$_{2}$Fe$_{17}$ are assumed to be the same to each other in Eq.~\eqref{eq:formationenergy}.
We also take account of the CeFe$_{2}$ phase,
another competing phase
which appears in the phase diagram as the next nearest phase of CeFe$_{12}$,
with calculation of the formation energy of CeFe$_{12}$ from CeFe$_{2}$:
\begin{equation}
  E[\text{CeFe}_{12}] - \left(E[\text{Ce}\text{Fe}_{2}] + 10E[\text{Fe}]\right).
  \label{eq:formationenergy2}
\end{equation}

In Fig.~\ref{fig:formationenergy_1-12_2-17_1-2}, the formation energies for CeFe$_{12}$ are plotted and 
the corresponding results for $R$Fe$_{12}$ ($R=$ Zr, Nd, and Sm) are shown for comparison.
Nd and Sm are typical rare-earths used in permanent magnet compounds, 
and Zr is recently suggested as a stabilizer of the ThMn$_{12}$-type structure.
The values for Ce$^{4+}$Fe$_{12}$ are significantly less than those for NdFe$_{12}$, SmFe$_{12}$ and ZrFe$_{12}$.
Although
the formation energy against Ce$^{4+}$Fe$_{2}$ is not negative,
it is also much less than the values of the other $R$Fe$_{12}$ systems against $R$Fe$_2$.
Providing that the CeFe$_2$ phase can be suppressed in a process, 
which is achieved in synthesis of Ce$_{2}$Fe$_{17}$ or Sm$_{2}$Fe$_{17}$ in practice,
our result suggest the potential of Ce$^{4+}$ in stabilizing the CeFe$_{12}$ phase. 

To analyze the stabilization by a tetravalent element systematically, 
we consider hypothetical $R$Fe$_{12}$ systems
for $R=$ Pr, Nd, Sm, Gd, Dy, Ho, Er, Tm and Lu,
with the valency of $R$ assumed to be tetravalent.
We show
the formation energy of $R$Fe$_{12}$ from $R_{2}$Fe$_{17}$ + bcc-Fe 
as a function of $r_{R}^{\text{voronoi}}$ in Fig.~\ref{fig:formationenergy_1-12_2-17}.
The results we have reported in Ref.~\onlinecite{HaFuKiMi2018} are also plotted in the figure for comparison. 
$r_{R}^{\text{voronoi}}$ is defined as the radius of a sphere whose volume is equal to the Voronoi cell volume of the $R$ site in $R$Fe$_{12}$,
which can be regarded as an estimation of the atomic radius of $R$.
The data points form two distinct curves, which is differentiated by the valency of the $R$ element.
Interestingly, Hf and Zr, which are tetravalent and non-rare-earth elements, also follow this trend. 
The formation energies become minima around $r_{R}^{\text{voronoi}} \approx$ 1.62 \AA\;, 
and the tetravalent elements tend to yield lower formation energy than the trivalent elements.
The valency of the rare-earth elements significantly affects the stability, and 
is important as well as the atomic size.


\begin{figure}[t]
  \includegraphics[width=\hsize]{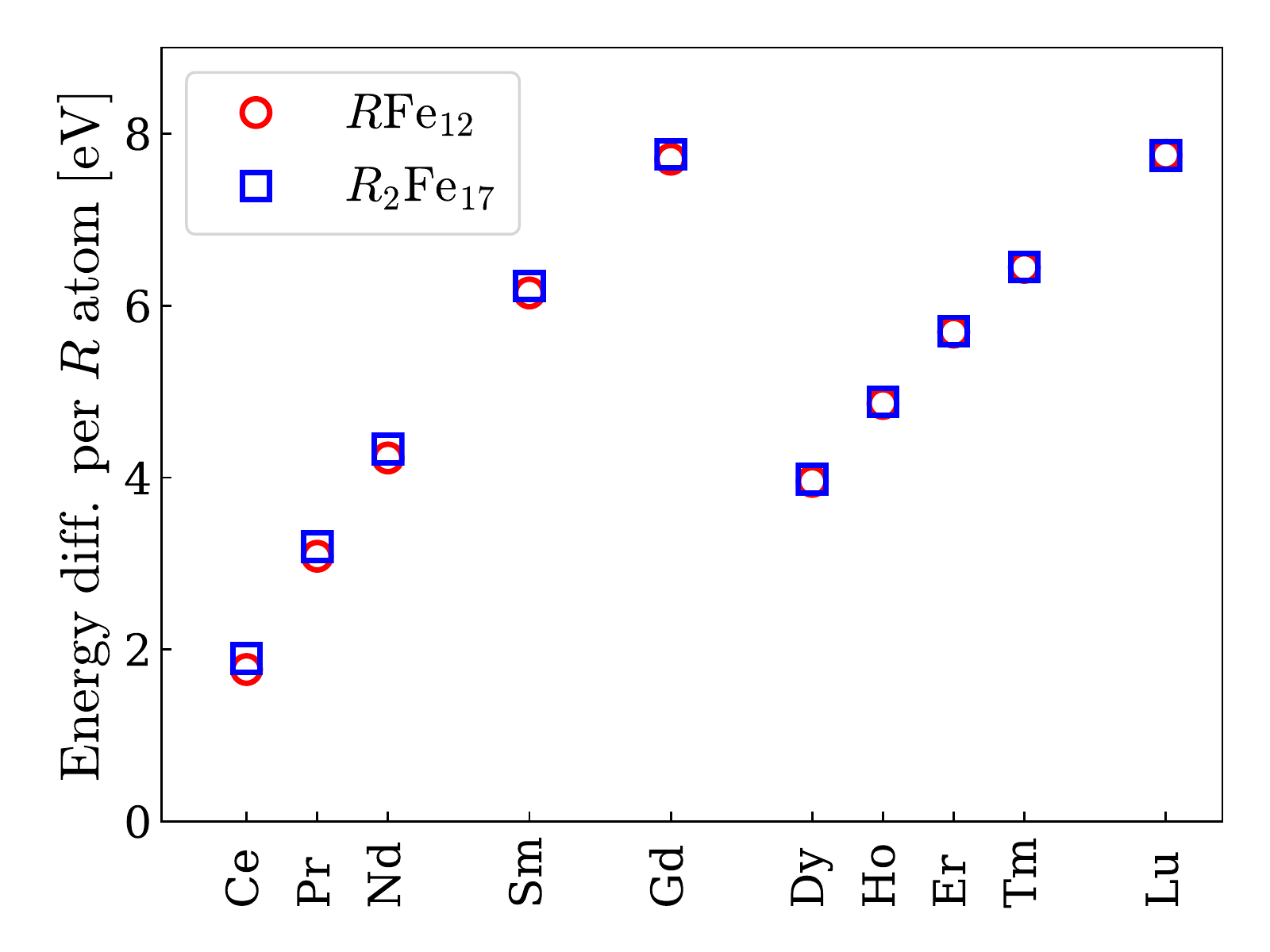}
  \caption{(Color online)
    Total energy difference per $R$ atom between $R^{3+}$Fe$_{12}$ and $R^{4+}$Fe$_{12}$
    ($E[R^{4+}\text{Fe}_{12}]-E[R^{3+}\text{Fe}_{12}]$),
    and that between $R^{3+}_2$Fe$_{17}$ and $R^{4+}_2$Fe$_{17}$
    (a half of $E[R^{4+}_2\text{Fe}_{17}]-E[R^{3+}_2\text{Fe}_{17}]$).
  }
  \label{fig:energydifference}
\end{figure}

We also investigate stability of the tetravalent phase over the trivalent phase.
Figure~\ref{fig:energydifference} shows the total energy difference per $R$ atom
between $R^{3+}$Fe$_{12}$ and $R^{4+}$Fe$_{12}$ (red) and
that between $R^{3+}_2$Fe$_{17}$ and $R^{4+}_2$Fe$_{17}$ (blue).
The values at $R=$ Ce are admittedly positive, which means that the tetravalent state is less stable than the trivalent state. 
This contradicts the experimental observation of Ce$_{2}$Fe$_{17}$ and
our expectation for CeFe$_{12}$. 
However, 
the energy difference systematically decreases as a function of the atomic number of $R$ decreases.
This means that there is a trend toward a tetravalent state when $R$ approaches Ce,
which is in agreement with the expectation and the experiment.
The deviation may be ascribed to theoretical errors coming from our calculational framework
using GGA and the open-core treatment, solution of which
is an open question at the present stage.

\begin{figure}[ht]
  \includegraphics[width=\hsize]{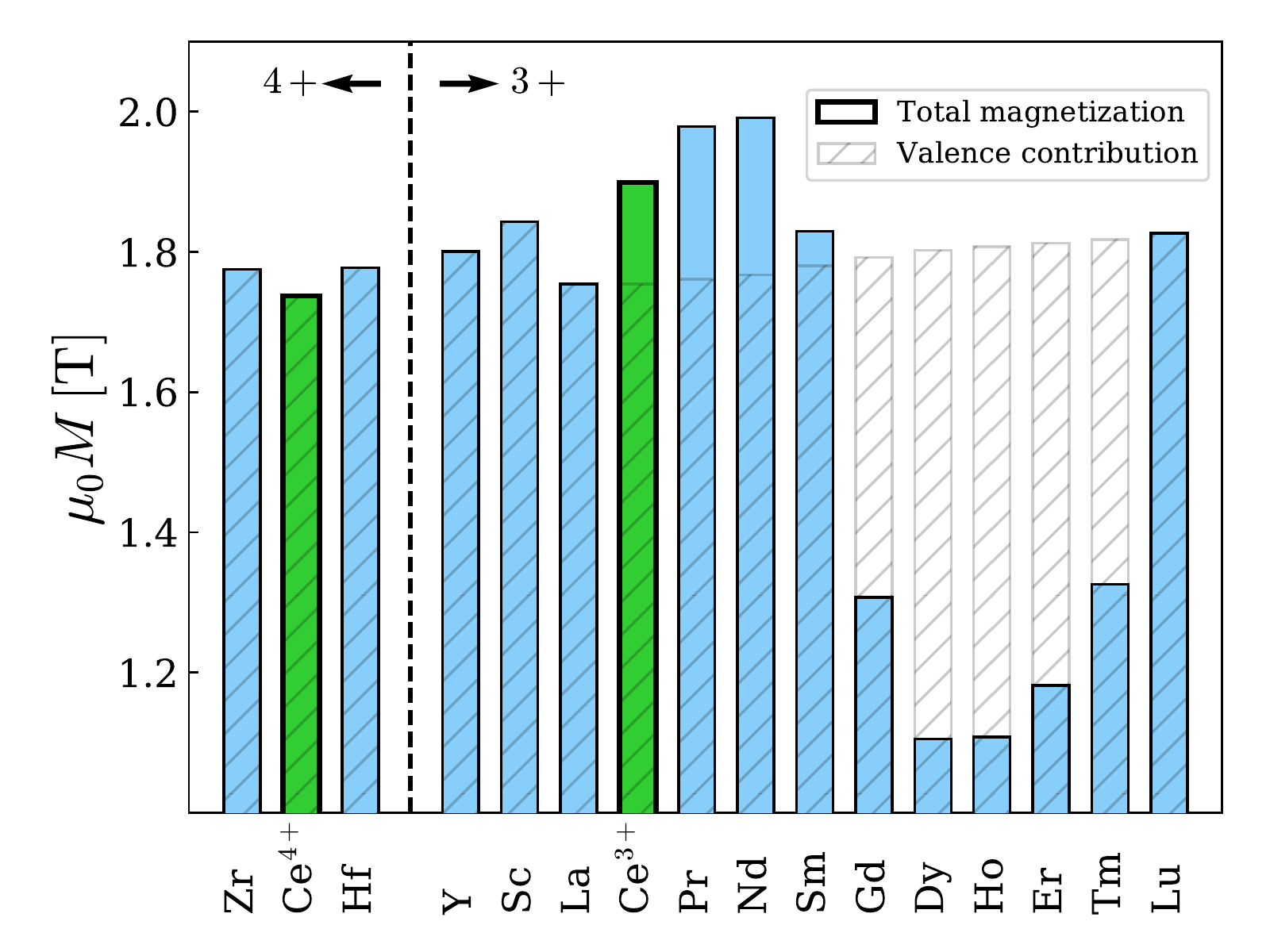}
  \caption{(Color online) Calculated magnetization of $R$Fe$_{12}$ in tesla.
    The filled bars indicate the total magnetization including the contribution from the 4$f$ electrons estimated by $g_{J}J$, 
    where $g_{J}$ denotes Lande g-factor and $J$ denotes the total angular momentum of the 4$f$ electrons.
    The shaded bars indicate the contribution from the valence electrons.
    The tetravalent elements, Zr, Ce$^{4+}$, and Hf are put on the left hand side and 
    the trivalent elements are on the right.}
  \label{fig:magnetization_1-12}
\end{figure}

Finally, we investigate the magnetization of CeFe$_{12}$.
Figure~\ref{fig:magnetization_1-12} shows the magnetizations of CeFe$_{12}$ and other $R$Fe$_{12}$ compounds.
All systems has a similar contribution to the magnetization from the valence electrons, and the difference comes mainly from 
the magnetization of the $f$-electrons.
Ce$^{4+}$Fe$_{12}$ has a lower magnetization than Ce$^{3+}$Fe$_{12}$,
large part of which is attributed to the local magnetic moment of Ce-4$f$ electrons.
Difference in hybridization between Ce and Fe also has a minor but discernible effect on the magnetization.


In conclusion, 
we performed first-principles calculation for CeFe$_{12}$ with the trivalent and tetravalent Ce.
The formation energies of CeFe$_{12}$ from Ce$_{2}$Fe$_{17}$ + bcc-Fe and CeFe$_{2}$ + bcc-Fe are calculated.
By comparison with the formation energies of $R$Fe$_{12}$ for $R=$ Nd, Sm, and Zr,  
it is found that the value for Ce$^{4+}$Fe$_{12}$ is lower than the other $R$Fe$_{12}$ systems. 
This result indicates that the tetravalent Ce is a promising stabilizer for the ThMn$_{12}$-type compounds. 
Our analysis revealed that the tetravalency of Ce is important in its functionality as a stabilizer.
We also calculated the magnetization of CeFe$_{12}$.
Ce$^{4+}$Fe$_{12}$ has a smaller magnetization than NdFe$_{12}$ and SmFe$_{12}$,
but its potential in stabilizing the ThMn$_{12}$ phase is still appealing.
We expect fractional doping of Ce can improve the stability of the $R$Fe$_{12}$ phase with a small loss of 
the magnetization. 
However, the amount of Ce 
is necessary to be tuned
to balance the magnetization and stability, 
which cannot be predicted by a simple interpolation in general even when the concentration
of the dopant is small,\cite{FuAkHaMi2018} in practical use.
Finding the optimum concentration of Ce is an important open problem, 
in which we should take account of other stabilizers.
We refer those who are interested in such a composition-optimization problem to Ref.~\onlinecite{FuHaHoMi2019},
one of our previous study with a machine learning technique.

We thank Kiyoyuki Terakura, Hisazumi Akai and Satoshi Hirosawa for fruitful discussion.
This work was supported by the Elements Strategy Initiative Project under the auspices of MEXT, and by MEXT as a social and scientific priority issue (Creation of new functional Devices and high-performance Materials to Support next-generation Industries; CDMSI) to be tackled by using the post-K computer. 
The computation was partly conducted using the facilities of the Supercomputer Center, the Institute for Solid State Physics, the University of Tokyo, and the supercomputer of ACCMS, Kyoto University. 
This research also used computational resources of the K computer provided by the RIKEN Advanced Institute for Computational Science through the HPCI System Research project (Project ID:hp180206).

\bibliography{./Reference}

\end{document}